\documentclass[nofootinbib,aps,a4paper,letterpaper,superscriptaddress,
twocolumn,times,eqsecnum]{revtex4}
\usepackage{amsmath}
\usepackage{amsfonts}
\usepackage{booktabs}
\usepackage{multirow}
\usepackage{siunitx} 
\usepackage{adjustbox}
\usepackage{microtype}
 \pdfoutput=1
\usepackage{graphicx}
\usepackage{color}
\usepackage{braket}
\usepackage{dcolumn}
\usepackage{bm,url}
\usepackage[linktocpage]{hyperref}
\usepackage{subfigure}
\usepackage{amsfonts}
\usepackage[usenames,dvipsnames,svgnames]{xcolor}  
\usepackage{hyperref}   
\definecolor{oxfordblue}{rgb}{0.0, 0.13, 0.28}
\definecolor{burgundy}{rgb}{0.5, 0.0, 0.13}
\definecolor{darkolivegreen}{rgb}{0.33, 0.42, 0.18}
\definecolor{darkblue}{rgb}{0,0,0.5}
\definecolor{richcarmine}{rgb}{0.84, 0.0, 0.25}
\definecolor{darkblue}{rgb}{0,0,0.5}
\definecolor{bluer}{rgb}{0.00,0.50,0.75}{}
\hypersetup{colorlinks=true, citecolor=red, linkcolor=blue,
 urlcolor = magenta, filecolor=magenta}

\begin{document}

\title{The generalized second law as a thermodynamic selection criterion for 
dynamical dark energy}

  \author{Marcos H. Cruz}
 \email{zs25021833@estudiantes.uv.mx}
   \affiliation{Facultad de F\'{\i}sica, Universidad Veracruzana 91097, 
Xalapa, Veracruz, M\'exico}
    
  \author{Miguel Cruz}
 \email{miguelcruz02@uv.mx}
  \affiliation{Facultad de F\'{\i}sica, Universidad Veracruzana 91097, 
Xalapa, Veracruz, M\'exico}
 
   \author{Joel 
Saavedra}
 \email{joel.saavedra@pucv.cl}
  \affiliation{Instituto de F\'\i sica, Pontificia Universidad Cat\'olica de 
Valpara\'\i 
so, Casilla 4950, Valpara\'\i so, Chile}
  
 \author{Emmanuel N. Saridakis}
 \email{msaridak@noa.gr}
 \affiliation{Institute for Astronomy, Astrophysics, Space Applications and 
Remote Sensing, National Observatory of Athens, 15236 Penteli, Greece}
 \affiliation{Departamento de Matem\'{a}ticas, Universidad Cat\'{o}lica del 
  Norte, Avda. Angamos 0610, Casilla 1280, Antofagasta, Chile}
 \affiliation{CAS Key Laboratory for Research in Galaxies and Cosmology, 
School 
  of Astronomy and Space Science,
  University of Science and Technology of China, Hefei 230026, China}
 
\begin{abstract}
The growing number of generalized horizon entropy proposals has led to a wide
variety of modified cosmological models, yet there is currently no general
physical principle capable of discriminating among them. We show at the level of an homogeneous background that the generalized second law (GSL) of thermodynamics provides such a criterion.
Considering a general modified Friedmann framework described by an arbitrary
function $f(H)$ of the Hubble parameter, and allowing the apparent horizon and
the cosmic fluid to evolve out of thermal equilibrium, we derive
model-independent constraints on the asymptotic scaling of the horizon entropy,
$S_A\propto A^k$. We find that phantom evolution requires
$k\ge(3\omega-1)/(2\omega)$, with $\omega$ the total fluid equation of state, 
whereas quintessence imposes the complementary
upper bound. Additionally, in the exact thermal equilibrium 
limit, the dynamical coupling strictly enforces $k \le 2$, recovering the 
non-equilibrium quintessence bound. The two bounds converge to the unique value 
$k=2$ as the phantom
divide is approached, indicating that a smooth crossing of the phantom divide
is thermodynamically associated with a quadratic entropy-area 
scaling, where the effective theory degenerates into a 
logarithmic gravity framework. Applying this criterion to representative generalized entropy models shows that
many commonly used proposals are constrained by the generalized second law,
whereas multiparameter constructions are naturally compatible with the required
asymptotic behavior. Finally, we demonstrate that the same thermodynamic
selection principle extends to derivative-dependent cosmologies described by
$f(H,\dot H)$, highlighting its robustness beyond entropy functionals that
depend only on the horizon area.
\end{abstract}

\maketitle

\section{Introduction}
\label{sec:intro}

The possible connection between gravity and thermodynamics is one of the deepest
conjectures of modern theoretical physics. The discovery that black holes 
possess a
temperature and an entropy \cite{Bekenstein1973,Hawking1975} established that 
the
laws of horizon mechanics are genuine thermodynamic laws, while Jacobson
demonstrated that Einstein's field equations can be recovered as an equation of
state by imposing the Clausius relation on local Rindler horizons
\cite{Jacobson1995,Padmanabhan:2003gd,Padmanabhan:2009vy}. Since then, the 
thermodynamic interpretation of gravity has
been developed in many directions \cite{Padmanabhan2010}, including approaches 
in
which gravity itself is regarded as an emergent entropic interaction
\cite{Verlinde2011}. In all these formulations, the Bekenstein-Hawking entropy,
proportional to the horizon area, plays a central role.

In cosmology, the corresponding causal boundary is the apparent horizon of the
Friedmann-Lemaître-Robertson-Walker (FLRW) spacetime. Cai and Kim showed that
applying the first law of thermodynamics to the apparent horizon reproduces the
Friedmann equations \cite{Cai_2005}, a framework that was subsequently extended 
to many modified theories of gravity
\cite{ 
 Wu:2007se,MohseniSadjadi:2007zq,Cai:2007bh,  
Akbar:2008vz, Cai:2008ys,Bamba:2009ay, 
Jamil:2009eb,Wang:2009zv,Cai:2009qf, 
Bamba:2010kf,Sadjadi:2010kp, Hendi:2010xr,
Karami:2012fu,Salako:2013gka,Cao:2013xy, Momeni:2015fyt, 
Chakraborty:2015wma,Bahamonde:2016cul, 
Zubair:2016bpi, Rudra:2020rhs,Pourhassan:2020jfu,Jawad:2020wlg, 
Tsilioukas:2025dmy, 
Lymperis:2025vup, Yang_2024}. Since the apparent horizon is a 
dynamical marginally trapped
surface, its surface gravity and temperature require a careful treatment, 
leading
to a consistent positive horizon temperature
\cite{Hayward_1998,Hayward_1999,helou2015dynamicscosmologicalapparenthorizon,
helou2015dynamicskindstrappinghorizons,PhysRevD.92.024001,Pourhassan:2020jfu} 
(see also \cite{faraoni2015cosmological,Sebastiani_2023,Cruz_2024}  for 
comprehensive
discussions of apparent-horizon thermodynamics).

At the same time, considerable effort has been devoted to understanding the
microscopic origin of horizon entropy. Although General Relativity reproduces 
the
Bekenstein-Hawking area law, numerous quantum-gravitational, statistical and
holographic arguments predict deviations from it. In particular, logarithmic 
corrections arise
naturally in loop quantum gravity and string theory
\cite{Cai_2008,Salehi_2018}, Tsallis statistics leads to the power-law entropy
$S\propto A^\delta$ \cite{Tsallis1988479,tsallis2009introduction}, Barrow's
proposal associates quantum-gravitational deformations of the horizon with
$S\propto A^{1+\Delta/2}$ \cite{Barrow_2020}, and Kaniadakis statistics produces
a hyperbolic-sine deformation
\cite{PhysRevE.66.056125,PhysRevE.72.036108,Abreu_2021,Cruz_kan}. More recently,
several generalized constructions have been proposed, including multiparameter
and singularity-free entropies
\cite{PhysRevD.109.103515,Nojiri_2022,Nojiri2022,
PhysRevD.105.044042,ODINTSOV2023101159}, exponential quantum-gravity inspired
forms \cite{Majhi_2017,Liu_2022}, and two-parameter entropy functionals
\cite{Leizerovich_2026,luciano2026}. Since every entropy-area relation 
generates
a corresponding modified Friedmann equation,  through 
the application of the gravity-thermodynamics conjecture, each
extension  leads a distinct cosmological model, leading to a rich literature 
\cite{Lymperis:2018iuz, Moradpour:2016rcy,Moradpour:2017ycq, 
Moradpour:2018ivi,MohseniSadjadi:2010nu, Bamba:2018zil,Nojiri:2019skr,  
Hernandez-Almada:2021rjs, 
 Asghari:2021bqa,  Nojiri:2021jxf,  Saridakis:2020lrg, 
Giardino:2020myz, Barrow:2020kug, Leon:2021wyx,
Lymperis:2021qty, Ghoshal:2021ief,Jusufi:2021fek,Jizba:2022bfz,Keskin:2021zct,  
Odintsov:2022qnn,   Luciano:2022ely,Dheepika:2022sio,  Luciano:2022knb,   
Nojiri:2022nmu,  Luciano:2022pzg,  Basilakos:2023kvk,
Lambiase:2023ryq, Mohammadi:2023lss, Salehi:2023gzv, Anand:2025cer,
Jizba:2023fkp, Luciano:2023roh,  Jizba:2024klq, Petronikolou:2024zcj, 
 Mondal:2024wno, Asghari:2024sbu, Karabat:2024trf,Anand:2025rjg}.

This proliferation of generalized entropy proposals naturally raises the
question of whether a general physical principle can discriminate among them.
Although many entropy functionals have been motivated by microscopic arguments
or confronted with cosmological observations, there is still no criterion that
selects them solely on the basis of their thermodynamic consistency.

Motivated by the generalized entropy proposed in
Ref.~\cite{CHAGOYA2025139415}, in which different entropy corrections dominate
different stages of the cosmic evolution, we investigate whether the
generalized second law (GSL) of thermodynamics itself can determine which 
entropy corrections are
thermodynamically admissible. 

To this end, we consider a general modified Friedmann framework described by an
arbitrary function $f(H)$ of the Hubble parameter. Allowing the apparent
horizon and the enclosed cosmic fluid to evolve out of thermal equilibrium, we
derive the GSL without assuming any specific modified gravity theory or entropy
functional. We show that it reduces to a simple constraint on the entropy-area
scaling,
$S_A\propto A^k$, where the exponent $k$ depends only on the total
equation-of-state parameter of the cosmic fluid. Phantom and quintessence
evolution impose complementary bounds on $k$, and the resulting relations 
provide  a simple
thermodynamic criterion for constraining generalized entropy models.

We further examine the opposite limit of exact thermal equilibrium between the
apparent horizon and the cosmic fluid, showing that the same asymptotic entropy
scaling is recovered. Finally, we extend the analysis to derivative-dependent
cosmologies described by $f(H,\dot H)$ and show that the thermodynamic
criterion remains valid even though the horizon entropy is no longer a function
only of the apparent-horizon area.

The paper is organized as follows, outlining a clear logical progression from geometric modifications to thermodynamic constraints. Section~\ref{sec:thermo} reviews the thermodynamics of the apparent horizon and derives the horizon entropy for a general $f(H)$ cosmology. Building on this correspondence, Section~\ref{sec:selection} formulates the generalized second law and derives the corresponding constraints on the entropy-area scaling, effectively providing a macroscopic thermodynamic filter. Section~\ref{sec:applications} applies these constraints to test the viability of representative generalized entropy models found in the literature, and compares them with the exact thermal-equilibrium case. Finally, Section~\ref{sec:conclusions} summarizes the main results and highlights the consequences for standard cosmological models, while Appendix~\ref{sec:genef} extends the analysis to $f(H,\dot H)$ cosmologies.

\section{Thermodynamics of the apparent horizon}
 \label{sec:thermo}

The thermodynamic selection criterion developed in this work relies on the intimate correspondence between the cosmological dynamics and the entropy associated with the apparent horizon. We therefore begin by reviewing the thermodynamics of a spatially flat Friedmann-Lemaître-Robertson-Walker (FLRW) universe and deriving the horizon entropy corresponding to a general modified 
Friedmann equation of the form $f(H)$. This construction establishes the framework that will be used in the following sections to formulate and apply the generalized second law.
 
\subsection{FLRW universe and apparent horizon}

The gravity-thermodynamics framework, originally 
inspired by black-hole thermodynamics, provides the geometric and thermodynamic 
ingredients that will be employed throughout the rest of the paper. 
Following \cite{Cai_2005,faraoni2015cosmological} we start with  the FLRW line 
element  
\begin{equation}
    ds^{2}=-dt^{2}+a^{2}(t)\left[\frac{dr^{2}}{1-kr^{2}}
    +r^{2}(d\theta^{2}+\sin^{2}\theta\,d\varphi^{2})\right],
    \label{eq:flrw}
\end{equation}
where $a(t)$ is the scale factor and $k$ is the spatial-curvature parameter. Introducing the areal radius
\begin{equation}
    R(t,r)=a(t)r,
\end{equation}
the metric can be written in the warped-product form
\begin{equation}
    ds^{2}=h_{ab}dx^{a}dx^{b}
    +R^{2}(d\theta^{2}+\sin^{2}\theta\,d\varphi^{2}),
\end{equation}
where $h_{ab}=\mathrm{diag}\!\left[-1,a^{2}(t)/(1-kr^{2})\right]$, with $a,b=(t,r)$.

The apparent horizon is defined as the marginally trapped surface satisfying
$h^{ab}\partial_{a}R\,\partial_{b}R=0$,
which yields the horizon radius
\begin{equation}
    R_{A}=\frac{1}{\sqrt{H^{2}+\frac{k}{a^{2}}}},
\end{equation}
with $H=\dot a/a$ the Hubble parameter. Since $R_A$ evolves with the cosmic 
expansion, the apparent horizon is a dynamical causal horizon, to which one may 
consistently assign thermodynamic quantities such as surface gravity, 
temperature and entropy \cite{Hayward_1998,Hayward_1999,Cai_2005}.

The corresponding surface gravity is
\begin{equation}
    \kappa=
    \frac{1}{2\sqrt{-h}}
    \partial_{a}
    \!\left(
    \sqrt{-h}\,
    h^{ab}\partial_{b}R
    \right),
\end{equation}
which explicitly becomes
\begin{equation}
    \kappa=
    -\frac{R}{2}
    \left(
    \dot H+2H^{2}+\frac{k}{a^{2}}
    \right).
\end{equation}
Specializing to the spatially flat case ($k=0$), where $R_A=H^{-1}$, and 
introducing the deceleration parameter
$q=-1-\dot H/H^{2}$, one obtains \cite{Cruz_2024}
\begin{equation}
    \kappa=
    -\frac{1}{2R_A}(1-q).
\end{equation}
In an expanding universe the apparent horizon is a past-inner trapping horizon, 
implying that the physical temperature is given by $T_A=-\kappa/(2\pi)$ rather 
than $\kappa/(2\pi)$ 
\cite{helou2015dynamicscosmologicalapparenthorizon,
helou2015dynamicskindstrappinghorizons,PhysRevD.92.024001}. Hence,
\begin{equation}
    T_A
    =
    -\frac{\kappa}{2\pi}
    =
    \frac{1}{2\pi R_A}
    \left(
    1-\frac{\dot R_A}{2HR_A}
    \right)
    =
    \frac{1-q}{4\pi R_A},
    \label{eq:temp}
\end{equation}
which reduces to the familiar de Sitter expression $T_A=H/(2\pi)$ for $q=-1$.

We assume that the  total material of the universe  is described by 
a perfect fluid with energy-momentum tensor
\begin{equation}
    T_{ab}=(\rho_{eff}+p_{eff})u_a u_b+pg_{ab},
\end{equation}
where $\rho_{eff}$ and $p_{eff}$ are respectively the effective energy density 
and pressure of the  total universe fluid, i.e. 
corresponding to the sum of matter, radiation and dark energy. Then 
the associated work density is  \cite{Cai_2005,faraoni2015cosmological} 
\begin{equation}
    W=-\frac12T=\frac12(\rho_{eff}-p_{eff}),
    \label{eq:work}
\end{equation}
where $T=h^{ab}T_{ab}$ denotes the normal two-dimensional trace. The total energy enclosed by the apparent horizon is
\begin{equation}
    E=\rho_{eff} V_A=\frac{4\pi}{3}\rho R_A^3,
\end{equation}
whose differential becomes
$dE=\frac{4\pi}{3}R_A^3\,d\rho_{eff}+4\pi\rho_{eff} R_A^2\,dR_A$.
Using the definition of the apparent horizon, this expression can be recast as
\begin{equation}
    dE=
    \frac{4\pi}{3}R_A^3\,d\rho_{eff}
    +2\pi R_A^2(\rho_{eff}+p_{eff})\,dR_A
    +W\,dV_A,
    \label{eq:de}
\end{equation}
which constitutes the unified first-law relation for an arbitrary FLRW 
cosmology \cite{Sebastiani_2023}. In the following subsection, this relation 
will be combined with the modified Friedmann equations to derive the 
corresponding horizon entropy.

\subsection{Modified Friedmann equations and horizon entropy}

Having established the thermodynamic framework of the apparent horizon, we now 
introduce a general class of cosmological models in which deviations from 
General Relativity are encoded in an arbitrary function of the Hubble parameter. Before proceeding, it is important to clarify the scope of this approach. The $f(H)$ formalism employed here serves as a phenomenological description strictly valid at the background cosmic level. Without a specific covariant action (such as a full Horndeski or $f(R)$ Lagrangian), one cannot obtain the perturbation equations required to analyze structure formation or to ensure perturbative stability. Nonetheless, our aim is to construct a macroscopic thermodynamic sieve; any fully covariant theory that attempts to reproduce this background kinematics must, as a necessary condition, satisfy the generalized second law constraints derived below. Restricting ourselves to a spatially flat universe, we consider the 
modified Friedmann equations
\begin{eqnarray}
    3f(H)&=&8\pi\rho_{eff},
    \label{eq:fried1}\\
    -\frac{\dot H}{H}f'(H)&=&8\pi(\rho_{eff}+p_{eff}),
    \label{eq:fried2}
\end{eqnarray}
where the prime denotes differentiation with respect to $H$.
This 
parametrization encompasses a broad class of modified cosmological scenarios, including higher-curvature and generalized gravity theories 
\cite{ARCINIEGA2020135242,ARCINIEGA2020135272,Erices_2019,Cisterna_2020,
ROMEROFIGUEROA2026117529,Valletta_2026}. Note that the standard scenario is recovered for the particular choice $f(H)=H^{2}$.

Combining the unified first-law relation (\ref{eq:de}) with the modified 
Friedmann equations (\ref{eq:fried1})-(\ref{eq:fried2}), we obtain
\begin{equation}
     dE
     =
     WdV_A
     -\frac12f'(H)R_A\,dR_A
     +\frac14f'(H)
     \frac{\dot R_A}{H^2R_A}
     dR_A.
\end{equation}
Using the horizon temperature (\ref{eq:temp}), this can be rewritten 
as
\begin{eqnarray}
dE
&=&
WdV_A
+\frac{f'(H)}{2H}
\left(
-1+\frac{\dot R_A}{2HR_A}
\right)
dR_A
\nonumber\\
&=&
WdV_A
+
T_A\,
\pi f'(H)R_A^2\,dR_A.
\end{eqnarray}
Since the last expression has precisely the form of the first law of 
thermodynamics,
$
dE=WdV_A+T_AdS_A$, it  allows us to identify the differential of the 
apparent-horizon entropy as
\begin{equation}
dS_A
=
\pi f'(H)R_A^2\,dR_A,
\label{eq:entro}
\end{equation}
thereby establishing the direct connection between the modified cosmological 
dynamics and the thermodynamic properties of the apparent horizon. From a physical standpoint, this mapping represents a structured exchange between the bulk geometry and the horizon boundary. A modification in the background geometry, dictated by $f(H)$, inherently alters the energy balance and the heat flow evaluated at the causal boundary of the universe. In order to preserve the unified first law of thermodynamics, the degrees of freedom on the apparent horizon, encoded in its entropy, must deviate from the standard Bekenstein-Hawking area law to properly absorb this modified energy flux.
Finally, note that in the standard case where $f(H)=H^{2}$ one finds $f'(H)=2H=2R_A^{-1},$
which   yields $dS_A=2\pi R_A\,dR_A$, whose integration gives $S_A=\frac{A}{4}$,
with $A=4\pi R_A^2$   the apparent-horizon area, and setting the integration 
constant to zero. Hence, the standard Bekenstein-Hawking 
entropy is naturally recovered.

Conversely, Eq.~(\ref{eq:entro}) also provides the inverse correspondence 
between a prescribed entropy-area relation and the cosmological dynamics. 
Assuming a generic entropy functional
\begin{equation}
S_A=g(A),
\end{equation}
one finds
\begin{equation}
f(A)
=
-16\pi
\int
\frac{g'(A)}{A^2}\,
dA,
\end{equation}
where the prime now denotes differentiation with respect to the area. Hence, 
in the particular case
$
g(A)=\frac{A}{4}$,
one recovers
$
f(A)=4\pi A^{-1}$
or equivalently
$
f(H)=H^2 $
which reproduces the standard Friedmann equations.

This one-to-one correspondence between the modified Friedmann function $f(H)$ and the horizon entropy constitutes the fundamental ingredient of the thermodynamic selection criterion that will be developed in the next section.

\section{Derivation of the thermodynamic selection criterion}
 \label{sec:selection}
 
In this section we derive a model-independent thermodynamic criterion that constrains the admissible form of the horizon entropy and, consequently, the underlying modified cosmological 
dynamics. Starting from the generalized second law (GSL) in a non-equilibrium 
description, where the apparent horizon and the cosmic fluid are allowed to 
evolve at different temperatures, we show that the requirement of 
non-decreasing 
total entropy can be translated into simple bounds on the entropy-area scaling. 
These bounds depend only on the cosmic equation-of-state parameter and provide 
a general thermodynamic filter applicable to broad classes of  generalized 
entropy models.

\subsection{Non-equilibrium thermodynamics and the generalized second law}
\label{subsec:nonequilibrium_gsl}

We now depart from the   assumption of local thermal equilibrium 
between the apparent horizon and the cosmic fluid enclosed within it. In this 
non-equilibrium setting, the geometric boundary and the matter sector are 
treated as distinct thermodynamic systems, characterized in general by 
different temperatures, $T_A$ and $T_{eff}$, respectively. The apparent-horizon 
temperature remains fixed by the geometric relation in Eq.~\eqref{eq:temp}, 
whereas the fluid temperature must follow from the local thermodynamic 
evolution of the matter sector.

For a barotropic    cosmic fluid  satisfying
$
p_{eff}=\omega\rho_{eff}$, with $\omega$ the effective equation-of-state 
parameter of the universe,
the entropy of the total fluid enclosed by the apparent horizon obeys the Gibbs 
relation
\begin{equation}
T_{eff} dS_{eff}=dE+pdV_A,
\end{equation}
where $
E=\rho_{eff} V_A$
is the total internal energy of the fluid. Taking the cosmic-time derivative 
and using the conservation equation
\begin{equation}
\label{conserveq}
\dot{\rho}_{eff}+3H\rho_{eff}(1+\omega)=0,
\end{equation}
together with the modified Friedmann equations (\ref{eq:fried1}) and 
(\ref{eq:fried2}), we obtain
\begin{equation}
\dot{E}+p_{eff}\dot{V}_A
=
\frac{3f(H)(1+\omega)q}{2H^2}.
\end{equation}
Therefore, by comparison, the corresponding  entropy production rate within the 
Universe is \begin{equation}
\dot{S}_{eff}
=
\frac{3f(H)(1+\omega)q}{2H^2T_{eff}}.
\label{entropyprod}
\end{equation}

On the other hand, differentiating the apparent-horizon entropy in 
Eq.~\eqref{eq:entro} with respect to cosmic time gives
\begin{equation}
\dot{S}_A
=
\pi\frac{f'(H)}{H^2}(1+q).
\end{equation}
The generalized second law requires the total entropy of the horizon-plus-fluid 
system to be non-decreasing, namely
\begin{equation}
\dot{S}_{\mathrm{tot}}
=
\dot{S}_A+\dot{S}_{eff}
\geq 0.
\end{equation}
Hence, for a general $f(H)$ cosmology, the GSL takes the form
\begin{equation}
\dot{S}_{\mathrm{tot}}
=
\pi\frac{f'(H)}{H^2}(1+q)
+
\frac{3f(H)(1+\omega)q}{2H^2T_{eff}}
\geq 0.
\label{eq:gsl}
\end{equation}
Note that a dependence on the deceleration parameter $q$ was also obtained in 
Ref.~\cite{PhysRevD.109.103515}, where the universe was treated as an open 
thermodynamic system and the cosmological evolution induces a matter flux 
across the apparent horizon.

As we observe, in the present non-equilibrium description, the 
independent fluid temperature $T_{eff}$ modifies the thermodynamic viability 
conditions. In particular, during an accelerating quintessence-like phase, 
where $-1<q<0$,
with
$
 \omega>-1$,
the universe interior entropy decreases, namely 
\begin{equation}
\dot{S}_{eff}<0.
\end{equation}
Hence, the horizon contribution must   compensate for this entropy 
loss, and thus equation~\eqref{eq:gsl}   implies
\begin{equation}
f'(H)
\geq
\frac{3f(H)(1+\omega)|q|}
{2\pi T_{eff}(1+q)}.
\end{equation}

To eliminate the explicit dependence on the a priori unknown fluid temperature 
and obtain a fully self-consistent formulation of the GSL, we model the cosmic 
bulk as an isentropic perfect fluid with conserved particle number. In 
relativistic thermodynamics, the local temperature then evolves according to
\begin{equation}
\frac{\dot{T}_{eff}}{T_{eff}}
=
-3H
\left(
\frac{\partial p_{eff}}{\partial\rho_{eff}}
\right)_n.
\end{equation}
For a barotropic fluid with constant equation-of-state parameter, one has
$
\left(
\frac{\partial p_{eff}}{\partial\rho_{eff}}
\right)_n
=
\omega$,
and therefore
\begin{equation}
T_{eff}(t)
=
T_0a(t)^{-3\omega},
\end{equation}
where $T_0$ is the present fluid temperature and the present scale factor is 
set to $a_0=1$.
Additionally, since the continuity equation (\ref{conserveq})   yields
$
\rho_{eff}(t)
=
\rho_0a(t)^{-3(1+\omega)}$,
we finally  obtain
\begin{equation}
T_{eff}(\rho_{eff})
=
T_0
\left(
\frac{\rho_{eff}}{\rho_0}
\right)^{\frac{\omega}{1+\omega}}.
\end{equation}
Furthermore, since the first modified Friedmann equation (\ref{eq:fried1}) 
gives 
$
\frac{\rho_{eff}}{\rho_0}
=
\frac{f(H)}{f_0}$,
where
$
f_0\equiv f(H_0)$, we find that   the matter temperature is determined entirely 
by the modified gravitational dynamics, namely 
\begin{equation}
T_{eff}(H)
=
T_0
\left[
\frac{f(H)}{f_0}
\right]^{\frac{\omega}{1+\omega}}.
\end{equation}
Finally, substituting this result into the matter-entropy production rate 
(\ref{entropyprod}) yields
\begin{equation}
\dot{S}_{eff}
=
\mathcal{K}
\frac{q}{H^2}
f(H)^{\frac{1}{1+\omega}},
\label{matterentropy2}
\end{equation}
where we have defined
\begin{equation}
\mathcal{K}
:=
\frac{3(1+\omega)
f_0^{\frac{\omega}{1+\omega}}}
{2T_0}.
\end{equation}

Hence, using (\ref{matterentropy2}) instead of (\ref{entropyprod}) in   the 
GSL, we can  re-write it entirely in terms of the kinematic variables 
and the function governing the modified cosmological dynamics, namely 
\begin{equation}
\dot{S}_{\mathrm{tot}}
=
\frac{\pi}{H^2}
f'(H)(1+q)
+
\frac{\mathcal{K}}{H^2}
qf(H)^{\frac{1}{1+\omega}}
\geq 0,
\end{equation}
and therefore GSL validity leads to  
\begin{equation}
\pi f'(H)(1+q)
\geq
-\mathcal{K}q
f(H)^{\frac{1}{1+\omega}}.
\label{eq:ineq}
\end{equation}

Equation~\eqref{eq:ineq} is the fundamental thermodynamic constraint on the 
modified Friedmann function. It shows that the admissibility of a cosmological 
model is governed by a non-linear balance between the evolution of the 
geometric sector, encoded in $f'(H)$, and the matter contribution, encoded in 
the power $f(H)^{1/(1+\omega)}$. In the next subsection, we translate this 
inequality into the apparent-horizon area and derive the corresponding 
constraint on the scaling of the horizon entropy.

\subsection{Area mapping and entropy scaling}
\label{subsec:area_{eff}apping}

The thermodynamic viability condition derived in Eq.~\eqref{eq:ineq} can be 
recast as a geometric constraint by expressing the modified Friedmann function 
in terms of the apparent-horizon area. For a spatially flat universe, this is 
simply  $
A=4\pi H^{-2}$.
Defining
\begin{equation}
F(A)\equiv f\bigl(H(A)\bigr),
\end{equation}
we thus obtain
\begin{equation}
f'(H)
=
\frac{dF}{dA}\frac{dA}{dH}
=
-\frac{A^{3/2}}{\sqrt{\pi}}F'(A),
\end{equation}
where   primes    denote  differentiation with respect to the argument. 
Substitution into Eq.~\eqref{eq:ineq} yields
\begin{equation}
\sqrt{\pi}A^{3/2}F'(A)(1+q)
\leq
\mathcal{K}qF(A)^{\frac{1}{1+\omega}},
\label{eq:master_area_ineq}
\end{equation}
which is  the area representation of the generalized second law.

We now relate this inequality to the scaling of the horizon entropy. 
We assume   a power-law entropy relation of the form 
\begin{equation}
S_A\propto A^k,
\end{equation}
which is known to be quite general. Rewriting 
Eq.~\eqref{eq:entro} in terms of $A$ gives
\begin{equation}
dS_A
=
-\frac{1}{16\pi}A^2F'(A),dA,
\end{equation}
 and thus  $dS_A\propto A^{k-1}dA$, and therefore
$
F'(A)\propto A^{k-3}$. Finally, integration gives
\begin{equation}
F(A)=\beta A^{k-2},
\qquad
F'(A)=\beta(k-2)A^{k-3},
\label{areamaping}
\end{equation}
where $\beta>0$. Note that the standard case $f(H)= H^2$ 
corresponds to $F(A)\propto A^{-1}$ and hence to $k=1$, recovering the 
Bekenstein-Hawking area law.

Substituting the power-law form of $F(A)$ into Eq.~\eqref{eq:master_area_ineq}, 
we find
\begin{equation}
\sqrt{\pi}\beta(k-2)(1+q)
A^{k-\frac{3}{2}}
\leq
\mathcal{K}q
\beta^{\frac{1}{1+\omega}}
A^{\frac{k-2}{1+\omega}}.
\label{eq:asymptotic_A}
\end{equation}
As we can see, this relation translates the GSL into a direct constraint on the 
entropy exponent $k$.

We first consider an asymptotic phantom-driven Big Rip, with $\omega<-1$. In 
this regime, $H\rightarrow\infty$, $A\rightarrow0$, $q<-1$, and 
$\mathcal{K}<0$. For the relevant range $k<2$, both coefficients in 
Eq.~\eqref{eq:asymptotic_A} are positive, and the inequality takes the 
asymptotic form
\begin{equation}
C_1A^{k-\frac{3}{2}}
\leq
C_2A^{\frac{k-2}{1+\omega}},
\qquad
C_1,C_2>0.
\end{equation}
For this condition to hold as $A\rightarrow0$, the exponent on the left-hand 
side must be greater than or equal to that on the right-hand side, namely
\begin{equation}
k-\frac{3}{2}
\geq
\frac{k-2}{1+\omega}.
\end{equation}
Solving for $k$ yields the phantom bound
\begin{equation}
k
\geq
\frac{3\omega-1}{2\omega}.
\label{eq:k_bound}
\end{equation}
  As we can see, the 
Bekenstein-Hawking value $k=1$ does not satisfy the asymptotic phantom 
condition, and thus a Big-Rip evolution requires a generalized entropy with a sufficiently large area exponent.

\begin{figure*}[htbp!]
\centering 
\includegraphics[width=1.39 \columnwidth]{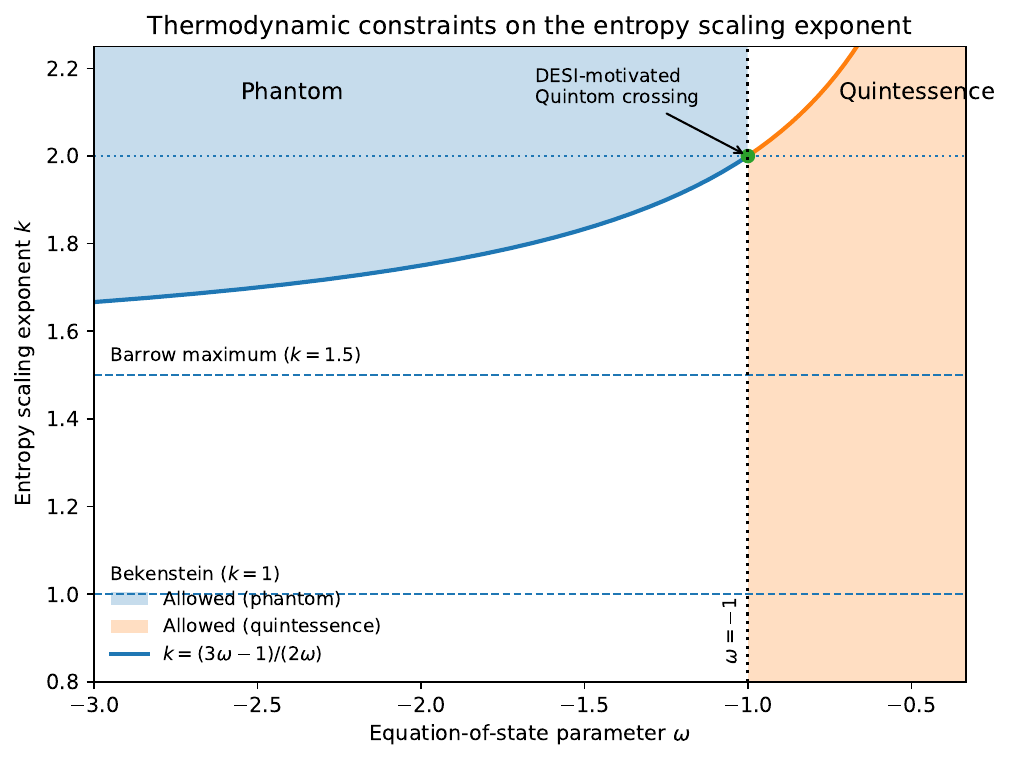}
 \caption{
{\it Thermodynamic constraints on the entropy scaling exponent $k$ as a 
function 
of the total equation-of-state parameter $\omega$, derived from the 
generalized second law (GSL). The solid curve corresponds to the critical value
$k=(3\omega-1)/(2\omega)$, which separates the thermodynamically allowed and 
forbidden regions. In the phantom regime ($\omega<-1$), viable cosmologies 
require $k$ to lie above the critical curve, whereas in the quintessence regime 
($-1<\omega<-1/3$), the allowed region lies below it. The horizontal dashed 
lines indicate the Bekenstein-Hawking entropy ($k=1$) and the maximum Barrow 
entropy exponent ($k=3/2$). The highlighted point at $(\omega,k)=(-1,2)$ marks 
the critical scaling associated with a smooth phantom crossing, illustrating 
that the generalized second law naturally selects an $A^2$ entropy behavior at 
the phantom divide.}}
  \label{fig:komega}
 \end{figure*}

We next consider the quintessence regime, $-1<\omega<-1/3$. In this case, 
$-1<q<0$, $\mathcal{K}>0$, and the decreasing Hubble parameter implies 
$A\rightarrow\infty$. For $k<2$, both sides of Eq.~\eqref{eq:asymptotic_A} are 
negative, so that multiplication by $-1$ gives
$
|C_1|A^{k-\frac{3}{2}}
\geq
|C_2|A^{\frac{k-2}{1+\omega}}$.
The large-area limit again requires
\begin{equation}
k-\frac{3}{2}
\geq
\frac{k-2}{1+\omega}.
\end{equation}
Since $\omega>-1$ but with $\omega<0$, isolating $k$ reverses the inequality 
and yields
\begin{equation}
k
\leq
\frac{3\omega-1}{2\omega}.
\label{eq:quintessence_bound}
\end{equation} 
As we observe, the quintessence case  is 
comfortably satisfied by the Bekenstein-Hawking value $k=1$.

Finally, the cosmological-constant limit, $\omega=-1$, forms the boundary 
between these regimes. At the exact de Sitter point, $\mathcal{K}=0$ and 
$q=-1$, 
so that
\begin{equation}
\dot{S}_{eff}=\dot{S}_A=\dot{S}_{\mathrm{tot}}=0.
\end{equation}
Hence, the GSL imposes complementary constraints in the two accelerating
regimes: phantom evolution requires a lower bound on $k$, whereas
quintessence evolution requires an upper bound.

The above thermodynamically
allowed regions are displayed in Fig.~\ref{fig:komega}. The two critical
branches converge to $k=2$ as $\omega\rightarrow-1$, while the exact de
Sitter solution constitutes a thermodynamic fixed point separating the
phantom and quintessence regimes. The implications of this convergence for
a dynamical crossing of the phantom divide are examined in the next
subsection.

\subsection{Crossing the phantom divide}
\label{subsec:desi_quintom}
Within the present thermodynamic framework, the crossing of
the phantom divide  
is particularly restrictive. The temperature relation derived under the 
adiabatic perfect-fluid assumption contains the factor $1+\omega$ in its 
exponent and must therefore be treated through the limiting behavior on the two 
sides of the divide. Having in mind the entropy-scaling constraints obtained 
in the previous subsection, namely 
$
k\geq\frac{3\omega-1}{2\omega}$ for the phantom phase and 
$
k\leq\frac{3\omega-1}{2\omega}$ for the quintessence phase, we deduce that a 
smooth transition through the phantom divide must approach both conditions 
continuously. Hence,
the lower bound inherited from the phantom side and the upper bound inherited 
from the quintessence side converge to the same value,
\begin{equation}
k\rightarrow2 .
\end{equation}
Consequently, thermodynamic consistency in the neighborhood of a smooth phantom 
crossing requires the apparent-horizon entropy to approach the quadratic area 
scaling $
S_A\propto A^2$.

Note, however, that this result should be distinguished from the thermodynamic 
behavior of an exact cosmological-constant solution. At the de Sitter fixed 
point we have 
$
\omega=-1$ and $q=-1$, and thus both the interior fluid and the horizon 
entropy-production rates vanish, namely 
$
\dot{S}_{eff}=
\dot{S}_A=0$,
and therefore
$
\dot{S}_{\mathrm{tot}}=0$.
At that   fixed point, the Bekenstein-Hawking entropy with $k=1$ is not 
excluded, because the generalized second law is saturated independently of the 
detailed entropy scaling. A phantom crossing, however, is not an exact de 
Sitter state but a genuinely dynamical passage through its neighborhood.  Thus, 
the quadratic scaling is selected by the continuity of the two thermodynamic 
bounds, rather than by the properties of the exact de Sitter point alone.

The same conclusion admits a direct interpretation in terms of 
the modified
Friedmann function, though it requires a careful treatment of the exact marginal 
case.
From the differential relation in Eq.~(\ref{eq:entro}), rewritten as
$dS_A = -\frac{1}{16\pi}A^2 F'(A)dA$, imposing the exact quadratic scaling
$S_A = \sigma A^2$ (which corresponds to $k=2$) directly yields
$
F'(A) = -\frac{32\pi\sigma}{A}$. Integration of this relation results in a logarithmic, rather than constant, behavior:
\begin{equation}
F(A) = -16\pi\sigma \ln(A^2) + C.
\end{equation}
Translating this back to the Hubble parameter using $A = 4\pi 
H^{-2}$, the corresponding
modified Friedmann function degenerates into a logarithmic gravity model, namely
\begin{equation}
f(H) = \gamma \ln H + c_0,
\end{equation}
where $\gamma = 64\pi\sigma$ and $c_0$ is an integration constant. Note that 
this
perfectly matches the equilibrium limit derived later in 
Section~\ref{subsec:exact_equilibrium},
where the boundary value $k=2$ corresponds to $n=0$, representing the exact 
marginal case
where the power law $f(H) \propto H^n$ degenerates into a logarithm.

In this exact logarithmic model, the derivative $f'(H) = 
\gamma/H$ is strictly non-zero
for any finite $H$. Consequently, a dynamical crossing of the phantom divide in 
a global
$S_A \propto A^2$ scenario does not occur via a vanishing geometric derivative. 
Instead,
the transition is driven purely by the cosmic kinematics coming to a momentary 
halt,
namely $\dot{H} \rightarrow 0$, which strictly defines $q \rightarrow -1$.

In this limit, evaluating the entropy production rates 
demonstrates thermodynamic consistency.
The horizon entropy production behaves as
\begin{equation}
\dot{S}_A \propto (1+q) \rightarrow 0.
\end{equation}
Simultaneously, since $\omega \rightarrow -1$, the prefactor in 
the matter entropy
production vanishes, causing $\dot{S}_{eff} \rightarrow 0$. Therefore, the total 
entropy
production approaches zero continuously, i.e.
\begin{equation}
\dot{S}_{\mathrm{tot}}(H_c)\rightarrow 0.
\end{equation}
It should be stressed that the logarithmic expression $f(H) = \gamma \ln H + c_0$ is not proposed as a full cosmological model, and it does not reproduce General Relativity in any limit. Physically, this is expected: in General Relativity, a smooth transition across the phantom divide is rigorously ruled out, as it would inevitably lead to catastrophic ghost instabilities. Therefore, the fundamental dynamics should not reduce to the standard one exactly at the crossing. The logarithmic modification acts as a transient, local effective description, a ``thermodynamic footprint'' of the new physics required to stabilize the transition. To understand how this fits into the global evolution of the Hubble profile, one could consider a composite scenario, such as $f(H) = H^2 + \gamma \ln(H/H_0)$. At high redshifts ($H \gg H_0$), the $H^2$ term dominates, successfully recovering General Relativity and the Bekenstein-Hawking entropy. However, as the universe expands and the Hubble parameter approaches the transition scale ($H \to H_c \approx H_0$), the kinematics momentarily halt ($\dot{H} \to 0$) and the geometric sector transiently degenerates into the logarithmic term, guiding the universe safely through the phantom divide without violating the generalized second law.

We conclude that there are two distinct mechanisms for a smooth 
phantom crossing.
In generalized multiparameter entropies, a dynamically varying effective 
exponent
$k_{\rm eff} \rightarrow 2$ can facilitate the crossing via a stationary 
geometric
sector ($f'(H_c)=0$). Conversely, for a global and exact $A^2$ scaling, the 
transition
is governed dynamically by the scale factor kinematics ($q=-1$). In both cases, 
the
generalized second law is fully respected.

\subsection{Comparison with regular cosmological models}
\label{subsec:comparison}
To further illustrate the predictive power of the derived thermodynamic criteria, it is instructive to contrast our findings with standard cosmological scenarios.
First, consider the $\Lambda$CDM limit. In the standard model, the Friedmann function is $f(H) = H^2 - \Lambda/3$, which perfectly corresponds to the Bekenstein-Hawking entropy ($k=1$). Because $\Lambda$CDM asymptotically approaches an exact de Sitter state with a strict equation of state $\omega = -1$, it converges to perfect thermodynamic equilibrium without ever crossing the phantom divide. Thus, it trivially satisfies the equilibrium bounds, and standard scenario remains thermodynamically safe indefinitely.
Second, for a regular quintessence scenario governed by a constant equation of state (e.g., $\omega = -0.8$), the upper bound requires $k \leq 2.125$. The standard framework with $k=1$ easily satisfies this condition and survives the future evolution without thermodynamic complications. However, a critical breakdown occurs if we consider a regular phantom fluid (e.g., $\omega = -1.2$). In this case, the lower phantom bound derived in Eq.~(\ref{eq:k_bound}) demands $k \geq 1.91$. This implies that the standard Bekenstein-Hawking entropy ($k=1$) is strictly forbidden. Physically, this demonstrates that a phantom universe governed by General Relativity will inevitably lead to a violation of the generalized second law as it approaches the Big Rip. To prevent this thermodynamic failure, the universe must adopt a modified geometry or a generalized form of entropy that increases the scaling exponent toward $k = 2$. This sharp contrast underscores the role of the generalized second law as a powerful theoretical filter for dynamical dark energy models.

\section{Application to generalized entropy models}
\label{sec:applications}

Having established the thermodynamic selection criterion, we now apply it to 
representative generalized entropy proposals that have been put forward in the 
literature. Our aim is to examine whether these models remain compatible with 
the generalized second law throughout the different stages of cosmic evolution 
and, in particular, whether they can consistently accommodate phantom evolution 
and a smooth  phantom-divide crossing. This analysis demonstrates the 
discriminating power of the proposed criterion and identifies the classes of 
generalized entropies that remain thermodynamically viable.

\subsection{Testing generalized entropy models}
\label{subsec:generalized_entropies}

The thermodynamic bounds derived in Sec.~\ref{subsec:area_{eff}apping} provide 
a direct criterion for testing proposed horizon entropies. In particular, the 
phantom regime requires the lower bound in Eq.~\eqref{eq:k_bound}, while 
quintessence imposes the complementary upper bound in 
Eq.~\eqref{eq:quintessence_bound}. For finite phantom equations of state, the 
critical exponent lies in the range $3/2<k<2$, approaching $k=2$ as 
$\omega\rightarrow-1$. We now examine several representative generalized 
entropy models according to their asymptotic area scaling.

\subsubsection{Barrow entropy}

Barrow entropy is motivated by a quantum-gravitational fractal deformation of 
the horizon and takes the form \cite{Barrow_2020}
\begin{equation}
S_B\propto A^{1+\Delta/2},
\qquad
0\leq\Delta\leq1.
\end{equation}
Its effective exponent is therefore $k=1+\Delta/2$. Even at maximal 
deformation, $\Delta=1$, one obtains only $k_{\rm max}=3/2$. Hence, Barrow 
entropy cannot satisfy the strict phantom bound and is thermodynamically 
incompatible with a finite phantom evolution within the present framework.

\subsubsection{Kaniadakis entropy}

Kaniadakis statistics provide a relativistically motivated extension of
Boltzmann-Gibbs thermodynamics
\cite{PhysRevE.66.056125,PhysRevE.72.036108,Abreu_2021}. Applied to the
apparent horizon, the entropy takes the form
\begin{equation}
S_K\propto\sinh(KA),
\end{equation}
where $K$ is the deformation parameter. The corresponding effective scaling 
exponent is
\begin{equation}
k_{\rm eff}
=
\frac{\partial\ln S_K}{\partial\ln A}
=
KA\,\coth(KA).
\end{equation}
Unlike the entropy models considered above, Kaniadakis entropy 
does not exhibit
an asymptotic power-law behavior of the form $S_A\propto A^k$, since its
effective exponent is area dependent. While $k_{\rm eff}\rightarrow1$
as $A\rightarrow0$, for large areas ($A\rightarrow\infty$) it grows without 
bound, $k_{\rm eff} \rightarrow \infty$. This extreme steepening strictly 
exceeds any finite upper bound imposed by the quintessence regime, effectively 
excluding it as a thermodynamically viable description for late-time 
accelerating expansion, unless $K=0$ exactly, in which case we recover standard Boltzmann-Gibbs thermodynamics.

\subsubsection{Tsallis entropy}

Tsallis non-extensive thermodynamics assumes 
\cite{Tsallis1988479,tsallis2009introduction}
\begin{equation}
S_T\propto A^\delta,
\end{equation}
so that the effective entropy exponent is simply $k=\delta$. Unlike the 
previous examples, Tsallis entropy can satisfy the phantom constraint, provided 
that
\begin{equation}
\delta
\geq
\frac{3\omega-1}{2\omega}.
\end{equation}
Thus, the generalized second law converts the non-extensivity index from a free 
parameter into a dynamically constrained quantity. In particular, the required 
value approaches $\delta\rightarrow2$ as the phantom divide is approached.

\subsubsection{Logarithmic quantum gravity entropy}

A logarithmic quantum-gravity entropy based on an exponential deformation has
been proposed in Refs.~\cite{Majhi_2017,Liu_2022},
\begin{equation}
S_{LQG}\propto e^{\beta A}-1.
\end{equation}
Unlike Kaniadakis entropy, this model exhibits well-defined asymptotic power-law
behavior. Assuming $\beta<0$, the entropy approaches a constant in the
large-area limit ($A\rightarrow\infty$), corresponding to an effective scaling
$k\rightarrow0$, which satisfies the quintessence bound. In the opposite limit
($A\rightarrow0$), the exponential can be expanded as
$
e^{\beta A}-1\simeq\beta A$,
recovering the linear scaling $S_{LQG}\propto A$ and therefore
$k\rightarrow1$. Within the thermodynamic selection criterion developed here,
this asymptotic behavior does not satisfy the phantom bound, nor does it allow
the entropy to approach the critical scaling $k=2$ associated with a smooth
crossing of the phantom divide.

\subsubsection{Extended entropic cosmology}

A generalized entropy containing logarithmic and power-law corrections was
proposed in Ref.~\cite{CHAGOYA2025139415},
\begin{equation}
S
=
\frac{A}{4}
+
\alpha\ln\left(\frac{A}{4}\right)
+
\sum_{j=0}^{N}
\sigma_j
\left(
\frac{A}{4}
\right)^{\frac{1+j}{2}},
\end{equation}
where $\alpha$ and $\sigma_j$ are free parameters. Its differential is
\begin{equation}
dS
=
\left[
\frac{1}{4}
+
\frac{\alpha}{A}
+
\sum_{j=0}^{N}
\widetilde{\sigma}_j
\left(
\frac{A}{4}
\right)^{\frac{j-1}{2}}
\right]dA,
\end{equation}
where $\widetilde{\sigma}_j$ absorbs the numerical factors arising from differentiation.

In the small-area limit, the dominant contribution depends on the coefficients.
For $\alpha\neq0$, the logarithmic correction controls the asymptotic behavior,
leading to $dS\propto A^{-1}dA$. If instead $\alpha=0$, the leading $j=0$
contribution gives $dS\propto A^{-1/2}dA$. In either case, the corresponding
asymptotic scaling lies below the phantom bound of
Eq.~\eqref{eq:k_bound}. Likewise, the inclusion of negative powers of the area,
although potentially relevant for early-universe applications, further lowers
the effective asymptotic scaling and therefore does not improve the situation
within the present thermodynamic framework.

In the large-area regime, the entropy is governed by the highest positive power,
\begin{equation}
S\propto A^{\frac{1+N}{2}},
\end{equation}
corresponding to
\begin{equation}
k=\frac{1+N}{2}.
\end{equation}
In particular, the choice $N=3$ yields the limiting scaling $k=2$ selected at
the phantom divide. Nevertheless, because the complete entropy contains several
competing contributions, the effective scaling generally evolves with the area.
Consequently, the thermodynamic implications of this model depend on which term
dominates in the cosmological regime under consideration, as well as on the
values of the free parameters.

\subsubsection{Luciano-Saridakis entropy}

The generalized entropy proposed by Luciano and Saridakis relaxes the usual
separability assumption and takes the two-power form
\cite{Leizerovich_2026,luciano2026}
\begin{equation}
S_{LS}
=
\gamma_1A^\delta
+
\gamma_2A^\epsilon,
\end{equation}
where $\gamma_1$, $\gamma_2$, $\delta$, and $\epsilon$ are constants. For
positive coefficients and distinct exponents, the small-area limit is governed
by
\begin{equation}
k_{\rm min}
=
\min(\delta,\epsilon),
\end{equation}
which should satisfy the lower bound in Eq.~\eqref{eq:k_bound}. Conversely, the
large-area limit is controlled by
\begin{equation}
k_{\rm max}
=
\max(\delta,\epsilon),
\end{equation}
which should satisfy Eq.~\eqref{eq:quintessence_bound}.

Unlike single-power entropies, the two asymptotic regimes are governed by
different exponents. This additional freedom allows the entropy to accommodate
independently the thermodynamic requirements associated with the small- and
large-area limits. In particular, the model naturally admits the possibility
that the effective scaling approaches the limiting value $k=2$ near the
phantom divide, while exhibiting different asymptotic behavior away from the
transition. It therefore constitutes a particularly interesting example within
the thermodynamic selection framework developed in the present work.
\subsubsection{Five-parameter generalized entropy}

Finally, we consider the multiparameter generalized entropy proposed as a
universal and singularity-free construction in
Refs.~\cite{PhysRevD.109.103515,Nojiri2022,Nojiri_2022,PhysRevD.105.044042,
ODINTSOV2023101159}. This class incorporates hyperbolic functions,
schematically of the form
\begin{equation}
\tanh\left(\alpha S^\beta\right),
\end{equation}
and possess sufficient parametric freedom for its effective scaling exponent
to vary across different cosmological regimes.

In the large-area limit, the hyperbolic function saturates and the effective
scaling approaches $k\rightarrow0$, remaining consistent with the
quintessence bound. In the opposite, small-area limit, the expansion
\begin{equation}
\tanh\left(\alpha S^\beta\right)
\simeq
\alpha S^\beta
\end{equation}
reduces the entropy to an effective power-law form. The corresponding
asymptotic exponent can therefore be directly confronted with the phantom bound
of relation \eqref{eq:k_bound}, providing constraints on the free parameters of 
the
model through their small-area behavior.

The comparison presented above reveals a common structural pattern. Entropies
characterized by a fixed asymptotic scaling generally satisfy the
thermodynamic requirements only within a restricted range of parameters, or in
some cases not at all. By contrast, multiparameter constructions possess
sufficient flexibility to exhibit different asymptotic scalings in the
small- and large-area regimes, making them particularly well suited for
comparison with the thermodynamic selection criterion developed in this work.
The thermodynamic classification of the generalized entropy models discussed
in this section is summarized in Table~\ref{tab:entropy_models}.

Although the above analysis has been carried out within a non-equilibrium
thermodynamic description, it is natural to ask whether the main conclusions
depend on this assumption. We therefore repeat the analysis in the opposite
limiting case of exact thermal equilibrium between the horizon and the cosmic
fluid.
\begin{table*}[t]
\centering
\caption{ 
Thermodynamic classification of the generalized horizon entropy models
considered in subsection \ref{subsec:generalized_entropies}. The phantom and 
quintessence columns refer,
respectively, to the asymptotic small-area ($A\rightarrow0$) and
large-area ($A\rightarrow\infty$) limits. A check mark indicates
compatibility with the corresponding GSL constraint, a cross denotes
incompatibility, ``Conditional'' indicates that viability depends
on the choice of model parameters, while ``N/A'' stands for non-applicable. The 
last column shows whether the
entropy can reproduce the critical scaling $S_A\propto A^2$ associated
with a smooth phantom transition.
} 
\label{tab:entropy_models}
\begin{ruledtabular}
\begin{tabular}{lcccc}
Entropy model
&
Effective asymptotic scaling
&
Phantom
&
Quintessence
&
Phantom-divide crossing
\\
Bekenstein-Hawking
&
$k=1$
&
$\times$
&
$\checkmark$
&
$\times$
\\
Barrow
&
$1\leq k\leq 3/2$
&
$\times$
&
$\checkmark$
&
$\times$
\\
Kaniadakis
&
$k_{\rm eff}
= 
KA\,\coth(KA)$
&
N/A
&
$\times$ (unless   $K=0$)
&
N/A
\\ 
Tsallis
&
$k=\delta$
&
Conditional
&
Conditional
&
Conditional
\\
Logarithmic quantum gravity
&
$k\rightarrow1$ for $A\rightarrow0$;
$k\rightarrow0$ for $A\rightarrow\infty$
&
$\times$
&
$\checkmark$
&
$\times$
\\
Extended entropic cosmology
&
Regime- and parameter-dependent
&
$\times$
&
Conditional
&
Conditional
\\
Luciano-Saridakis
&
$k_{\rm min}=\min(\delta,\epsilon)$;
$k_{\rm max}=\max(\delta,\epsilon)$
&
Conditional
&
Conditional
&
Conditional
\\
Five-parameter generalized entropy
&
Dynamical effective exponent
&
Conditional
&
$\checkmark$
&
Conditional
\end{tabular}
\end{ruledtabular}
\end{table*}

\subsection{Thermodynamic equilibrium}
\label{subsec:exact_equilibrium}

The thermodynamic selection criterion derived above was obtained within a
non-equilibrium description, where the apparent horizon and the cosmic fluid
may have different temperatures. It is therefore natural to ask whether the
main conclusions depend on this assumption. To address this question, we
consider the opposite limiting case of exact local thermal equilibrium, namely
\begin{equation}
T_m=T_A,
\end{equation}
which may provide a reasonable approximation during sufficiently late stages of cosmic evolution \cite{Mimoso_2016}. Using the apparent-horizon
temperature,
\begin{equation}
T_A=\frac{H}{4\pi}(1-q),
\end{equation}
the generalized second law, Eq.~\eqref{eq:gsl}, becomes
\begin{equation}
H\frac{f'(H)}{f(H)}(1+q)
+
\frac{6(1+\omega)q}{1-q}
\geq0.
\label{eq:gsl_eq}
\end{equation}
For a power-law modified Friedmann function, $f(H)\propto H^n$, 
equation \eqref{eq:gsl_eq} reduces to
\begin{equation}
n(1+q) \geq -\frac{6(1+\omega)q}{1-q}.
\label{eq:n_equilibrium}
\end{equation}
Crucially, the purely kinematic parameter $q$ and the fluid 
parameter $\omega$ cannot be treated as independent variables. The modified 
Friedmann equations impose a strict dynamical coupling between the geometric 
sector and the cosmic fluid. From the ratio of equations \eqref{eq:fried1} and 
\eqref{eq:fried2}, one obtains the exact relation 
\begin{equation}
n(1+q) = 3(1+\omega).
\label{eq:n_coupling}
\end{equation}
Substituting this coupling into the equilibrium GSL condition 
yields 
\begin{equation}
n(1+q) \left[ 1 + \frac{2q}{1-q} \right] = n \frac{(1+q)^2}{1-q} \geq 0.
\end{equation}
In any relevant expanding accelerating regime (quintessence 
$-1<q<0$, or phantom $q<-1$), the denominator $1-q$ is strictly positive, and 
the squared term $(1+q)^2$ is non-negative. Therefore, thermodynamic consistency 
directly demands:
\begin{equation}
n \geq 0.
\end{equation}
Using the area mapping established previously, $f(H)\propto 
H^{4-2k}$, the geometric exponent is related to the entropy exponent through 
$n=4-2k$. Therefore, the exact equilibrium analysis yields the strict upper 
bound:
\begin{equation}
k \leq 2.
\end{equation}
This result is highly significant. Rather than providing an 
independent lower bound, the strict thermal equilibrium limit exactly recovers 
and converges with the quintessence upper bound derived from the non-equilibrium 
analysis. The confluence of both thermodynamic descriptions demonstrates that a 
smooth crossing of the phantom divide inevitably traps the exponent, forcing it 
globally into the marginal limit $k=2$.

\section{Conclusions}
\label{sec:conclusions}

The large variety of generalized entropy-area relations proposed in recent years
has generated an equally broad class of modified cosmological models, yet there
is still no general principle capable of discriminating among them. In this work
we have shown within the context of an homogeneous background that the generalized second law (GSL) of thermodynamics provides
such a criterion. Working within a general modified Friedmann framework,
described by an arbitrary function $f(H)$, and allowing the apparent horizon and
the cosmic fluid to evolve out of thermal equilibrium, we found that the
thermodynamic consistency of the cosmological evolution reduces to a simple
constraint on the entropy-area scaling 
\begin{equation}
S_A\propto A^k .
\end{equation}
The allowed values of the exponent $k$ depend only on the total
equation-of-state parameter of the cosmic fluid. In particular, phantom
evolution requires
$k\ge (3\omega-1)/(2\omega)$, whereas accelerating quintessence requires the
complementary upper bound. As the phantom divide $\omega=-1$ is approached, the
two bounds converge to the unique limiting value
\begin{equation}
k=2,
\end{equation}
indicating that a smooth crossing of the phantom divide is thermodynamically
associated with a quadratic entropy-area scaling.

Applying this criterion to representative generalized entropy models reveals a
clear hierarchy. Standard Bekenstein-Hawking and Barrow entropies do not satisfy
the phantom bound, while Tsallis entropy remains compatible only within a
restricted parameter range. In contrast, multiparameter constructions, such as
the Luciano-Saridakis entropy, possess sufficient freedom to realize different
effective entropy scalings in different cosmological regimes and can naturally
accommodate the limiting behavior required near the phantom divide. Applying the
same analysis to the generalized entropy of Ref.~\cite{CHAGOYA2025139415}
further shows that the dominant entropy corrections in different cosmological
epochs must satisfy the corresponding thermodynamic bounds.

As an independent consistency check, we repeated the analysis 
assuming exact thermal equilibrium between the apparent horizon and the cosmic 
fluid. By enforcing the strict dynamical coupling between the geometric sector 
and the cosmic fluid, we found that the generalized second law enforces the 
global bound $k \le 2$. The convergence of this equilibrium limit with the 
non-equilibrium quintessence bound reveals a profound thermodynamic saturation: 
a smooth transition across the phantom divide inevitably traps the entropy-area 
exponent exactly at the limiting value $k=2$.

We further extended the analysis to derivative-dependent cosmologies described
by a generalized function $f(H,\dot H)$. In this broader framework the horizon
entropy is no longer determined solely by the apparent-horizon area, but
acquires an explicit dependence on higher-order cosmic kinematics. Nevertheless,
the generalized second law continues to impose the same qualitative
thermodynamic restrictions. In particular, a smooth crossing of the phantom
divide requires the effective geometric sector to become momentarily stationary,
corresponding to $\dot{H} = 0$ in the $f(H)$ formulation and 
$\dot f\to0$ in the
generalized $f(H,\dot H)$ framework.

In summary, our results indicate that the generalized second law can be viewed 
not
only as a consistency condition but also as a useful theoretical tool for
constraining generalized horizon entropies and the corresponding modified
cosmological models. In this sense, horizon thermodynamics provides guidance
that is complementary to microscopic constructions and cosmological
observations.  While a 
smooth transition from a phantom to a quintessence regime is structurally 
forbidden for a canonical and minimally coupled single scalar field in standard 
General Relativity, often necessitating highly fine-tuned $k$-essence models to 
circumvent ghost instabilities \cite{Nojiri2026}, generalized entropies and 
modified gravity theories can naturally accommodate such dynamics. Our findings 
demonstrate that within these modified frameworks, the phantom crossing is not 
an \textit{ad hoc} feature requiring artificial fine-tuning, but rather a 
physically constrained transition governed by the generalized second law. 
Consequently, the generalized second law acts as a crucial theoretical sieve, 
ensuring that the alternative cosmological models capable of reproducing the 
latest observational data also remain fundamentally sound from a macroscopic 
thermodynamic perspective.

Finally, the present framework opens direct avenues for 
observational confrontation. Future work will map the derived thermodynamic 
windows onto the precise phenomenological trajectories favored by the latest 
cosmological surveys, such as the DESI DR2 results, which exhibit a statistical 
preference for an evolving equation of state crossing the phantom divide at late 
times, as well as constraints from future gravitational wave detectors \cite{Yang_2024, Wang_2020}. Imposing this thermodynamic sieve, combined with dynamic stability 
criteria (such as the absence of ghost instabilities), could significantly 
restrict the parameter space of viable generalized gravity models.

\begin{acknowledgments}
M.~Cruz's work was partially supported by S.N.I.I. (SECIHTI-M\'exico). J.~Saavedra acknowledges the FONDECYT grant N°1220065, Chile.  E. N. Saridakis acknowledges the contribution of the LISA CosWG, and COST Actions CA21106 ``COSMIC WISPers in the Dark Universe: Theory, astrophysics and experiments'', CA21136 
``Addressing observational tensions in cosmology with 
  systematics and fundamental physics (CosmoVerse)'',    CA23130 
``Bridging high and low energies in search of quantum gravity (BridgeQG)'', and CA24101 ``Testing Fundamental Physics with Seismology''.
\end{acknowledgments}
\appendix
\section{Generalized framework: $f(H,\dot{H})$}
\label{sec:genef}

In the main text we considered modified cosmological dynamics described by a
geometric function depending only on the Hubble parameter, $f(H)$. Here we
extend the analysis to the more general class of derivative-dependent theories,
in which the gravitational sector depends explicitly on both $H$ and its first
time derivative, $\dot H$. We derive the corresponding horizon entropy and
generalized second law, and examine how the thermodynamic selection criterion
is modified in this broader framework.

\subsection{Horizon entropy in derivative-dependent cosmologies}

We consider the generalized cosmological framework in which the effective
gravitational sector depends on both the Hubble parameter and its first
derivative, namely $f(H,\dot H)$. In such theories the horizon entropy is no
longer expected to depend only on the apparent-horizon area. Indeed, according
to Wald's Noether-charge formalism
\cite{Wald_1993,Iyer_1994}, the entropy of any
diffeomorphism-invariant gravitational theory is determined by the
gravitational Lagrangian itself. Since, in a flat FLRW spacetime,
\begin{equation}
R=6\left(2H^2+\dot H\right),
\end{equation}
the entropy generally acquires an explicit dependence on the cosmic
kinematics.

The generalized first Friedmann equation is written as
\begin{equation}
3f(H,\dot H)=8\pi\rho_{eff},
\label{eq:gen_friedmann1}
\end{equation}
where $f(H,\dot H)$ is an arbitrary function of $H$ and $\dot H$. Assuming the
standard conservation law (\ref{conserveq}), differentiation of
Eq.~\eqref{eq:gen_friedmann1} gives
\begin{equation}
8\pi\dot\rho_{eff}=3\dot f,
\label{eq:chain_rule}
\end{equation}
with
\begin{equation}
\dot f
=
\frac{\partial f}{\partial H}\dot H
+
\frac{\partial f}{\partial\dot H}\ddot H.
\end{equation}
Using the conservation equation, one obtains the generalized second Friedmann
equation
\begin{equation}
8\pi(\rho_{eff}+p_{eff})
=
-\frac{\dot f}{H},
\label{eq:gen_friedmann3}
\end{equation}
which reduces to the $f(H)$ case when
$\partial f/\partial\dot H=0$.

The unified first law, Eq.~(\ref{eq:de}), remains valid independently of the
particular gravitational dynamics. Substituting
$d\rho_{eff}=(3/8\pi)\dot f\,dt$
together with Eq.~\eqref{eq:gen_friedmann3}, the heat flow through the apparent
horizon becomes
\begin{equation}
T_A\,dS_A
=
\frac12R_A^3\dot f\,dt
-
\frac14R_A^3\dot f\,dR_A.
\label{eq:heat_flux1}
\end{equation}
Using
$dt=dR_A/\dot R_A$
and the horizon temperature (\ref{eq:temp}), this reduces to
\begin{equation}
dS_A
=
-\pi R_A^2
\frac{\dot f}{\dot H}
\,dR_A.
\label{eq:entropy_diff_raw}
\end{equation}
Substituting the explicit expression for $\dot f$ finally yields
\begin{equation}
dS_A
=
\pi R_A^2
\left(
\frac{\partial f}{\partial H}
+
\frac{\partial f}{\partial\dot H}
\frac{\ddot H}{\dot H}
\right)
dR_A.
\label{eq:entropy_diff_final}
\end{equation}
Equation~\eqref{eq:entropy_diff_final} generalizes the entropy differential
derived in Sec. \ref{sec:thermo}. Unlike the $f(H)$ case, the horizon entropy 
is 
no longer a
function only of the apparent-horizon area, but depends explicitly on the
cosmic jerk through the ratio $\ddot H/\dot H$. Derivative-dependent theories
therefore incorporate the expansion kinematics directly into the
thermodynamic description of the apparent horizon.

\subsection{Generalized second law and thermodynamic implications}

The entropy differential derived above allows the generalized second law to be
constructed following the same procedure as in the main text. Differentiating
Eq.~\eqref{eq:entropy_diff_raw} with respect to cosmic time, the horizon entropy
production rate becomes
\begin{equation}
\dot S_A=-\pi R_A^4\dot f.
\end{equation}
Using the Gibbs relation for the cosmic fluid together with the generalized
Friedmann equations, the total entropy production is
\begin{equation}
\dot S_{\rm tot}
=
-\frac{\dot f}{H^3}
\left(
\frac{\pi}{H}
+
\frac{q}{2T_{eff}}
\right)
\ge0,
\label{eq:gsl_master}
\end{equation}
which is the natural extension of the generalized second law to
$f(H,\dot H)$ cosmologies.

In the limiting case of local thermal equilibrium,
$T_{eff}=T_A$,
Eq.~\eqref{eq:gsl_master} reduces to
\begin{equation}
\dot S_{\rm tot}
=
-\frac{\pi}{H^4}
\dot f
\left(
\frac{1+q}{1-q}
\right)
\ge0.
\label{eq:gsl_equilibrium}
\end{equation}
Since $\pi/H^4>0$ in an expanding universe, thermodynamic consistency requires
$\dot f\le0$ during quintessence evolution ($-1<q<0$), whereas the sign
reverses in the phantom regime ($q<-1$), implying $\dot f\ge0$.

Away from equilibrium, the fluid temperature follows
\begin{equation}
T_{eff}
=
T_0
\left(
\frac{f}{f_0}
\right)^{\frac{\omega}{1+\omega}},
\end{equation}
leading to
\begin{equation}
\dot S_{\rm tot}
=
-\frac{\pi}{H^4}\dot f
-
\frac{\mathcal K q}{H^3}
\frac{\dot f}
{f^{\frac{\omega}{1+\omega}}}
\ge0,
\label{eq:gsl_noneq}
\end{equation}
where $\mathcal K$ is the constant introduced in Sec. \ref{sec:selection}. As 
in 
the
$f(H)$ framework, the generalized second law imposes a non-linear constraint on
the evolution of the effective gravitational sector.

Finally, let us consider a smooth crossing of the phantom divide,
$\omega\rightarrow-1$. Although the exponent
$\omega/(1+\omega)$ in Eq.~\eqref{eq:gsl_noneq} becomes singular, the same limit
forces $(1+\omega)\rightarrow0$, and therefore
$\dot\rho\rightarrow0$ through the conservation equation. Using
Eq.~\eqref{eq:chain_rule}, this implies
\begin{equation}
\dot f
=
\frac{\partial f}{\partial H}\dot H
+
\frac{\partial f}{\partial\dot H}\ddot H
\longrightarrow0.
\label{eq:phantom_lock}
\end{equation}
Consequently,
\begin{equation}
\dot S_{\rm tot}\longrightarrow0,
\end{equation}
showing that the entropy production remains regular at the crossing. Therefore,
although derivative-dependent theories enrich the horizon entropy through
additional kinematic contributions, they preserve the same thermodynamic
selection principle obtained in the main text. In particular, a smooth crossing
of the phantom divide requires the effective gravitational sector to become
momentarily stationary, demonstrating that the thermodynamic criterion derived
for $f(H)$ cosmologies remains valid in the broader class of
$f(H,\dot H)$ theories.

\bibliography{refs}
\end{document}